\documentclass{elsart}

\usepackage{graphicx}
\usepackage{amssymb}
\usepackage{lineno}
\usepackage{amsmath}

\newcommand{\tr}{\mbox{Tr}}

\newcommand{\beq}{\begin{equation}}
\newcommand{\eeq}{\end{equation}}
\newcommand{\bea}{\begin{eqnarray}}
\newcommand{\eea}{\end{eqnarray}}

\begin{document}

\begin{frontmatter}
\title{Quantum communication through anisotropic Heisenberg XY spin chains}
\author[SIUCPhys,Beijing]{Z.-M. Wang}
\ead{mingmoon78@126.com}
\author[SIUCPhys,SIUCCS]{M. S. Byrd}
\ead{mbyrd@physics.siu.edu}
\author[Beijing]{B. Shao}
\author[Beijing]{J. Zou}
\address[SIUCPhys]{Department of Physics, Southern Illinois University, Carbondale, Illinois
  62901-4401}
\address[Beijing]{Department of Physics, Beijing Institute of Technology,
Beijing, 100081}
\address[SIUCCS]{Department of Computer Science, Southern Illinois University, Carbondale, Illinois 62901-4401}

\begin{abstract}
We study quantum communication through an anisotropic Heisenberg XY chain in
a transverse magnetic field. We find that for some time $t$ and anisotropy
parameter $\gamma$, one can transfer a state with a relatively high
fidelity. In the
strong-field regime, the anisotropy does not significantly affect the fidelity while in
the weak-field regime the affect is quite pronounced. The most
interesting case is the the intermediate regime where the
oscillation of the fidelity with time is low and the high-fidelity
peaks are relatively broad.  This would, in principle, allow for quantum
communication in realistic circumstances. Moreover, we calculate the
purity, or tangle, as a measure of the entanglement between one spin and all the
other spins in the chain and find that the stronger the anisotropy and
exchange interaction, the more entanglement will be generated for a
given time.
\end{abstract}

\maketitle

\begin{keyword}
Quantum State Transfer, Entanglement
\PACS 03.67Hk,03.65Ud,75.10Jm
\end{keyword}

\end{frontmatter}


\section{Introduction}

The transfer of a quantum state from one place to another is an important
task in quantum information processing. A quantum state prepared by one
party needs to be measured by another party at a distance. Long distance
communication between two parties, for example, in quantum key distribution
\cite{Ekert:91}, can be realized by means of photons. In this case, photons
have the advantage that they have an extremely small interaction with the
environment and also travel long distances quickly through optical fibers or
empty space. However, for short distance communication, such as connecting
distinct quantum processors or registers inside a
quantum computer \cite{Kielpinski/etal:02,Blinov/etal:04}, conditions
and requirements are different \cite{Bose:07}. Recently, quantum
communication through spin chains has been intensively investigated for this
purpose \cite{Bose:03,Christandl/etal:04,Christandl/etal:05}.

The primary scheme is that one quantum state is produced at one end of the
chain; it evolves naturally under spin chain dynamics; and at some time
$t$ we receive the state at the other end \cite{Bose:03}. For more
complex systems, Christandl et al. suggested a perfect state transfer
algorithm which can transfer an arbitrary quantum state between two ends of
a spin chain \cite{Christandl/etal:04}, or a more complex spin network \cite%
{Christandl/etal:05}. In addition, researchers have investigated
measurement-assisted optimal quantum communication by a single chain \cite%
{Burgarth/etal:06} and parallel chain
\cite{DBurgarth/etal:05,Burgarth/Bose:06}, the entanglement transfer through a
Heisenberg XY chain \cite{Subrahmanyam:04,Subrahmanyam/Lakshminarayan:06}
and parallel spin chains \cite{Wang/etal:07}, enhancement of state transfer
with energy current \cite{Wang/etal:08} and entanglement transfer by phase
control \cite{Maruyama:07,ZMWang/etal:07}. However, due to
the complexity of the problem researchers usually consider cases where the
magnetization (\textit{z} component of the total spin $\sum_{i}S_{i}^{z}$)
is a conserved quantity, which means that $[\sum_{i}S_{i}^{z},H]=0$. For the
cases that the Hamiltonian does not satisfy above conditions, L. Amico et
al. studied the dynamics of entanglement and found that the anisotropy of
the Hamiltonian has an evident effect on the evolution of entanglement
\cite{Amico/etal:04,LAmico/etal:04}. They
calculated the entanglement using the out-of-equilibrium correlation
functions, but did not investigate state transfer in these systems.
The anisotropy and magnetic
field effects on the entanglement transfer in two parallel Heisenberg spin
chains was later investigated in Ref.~\cite{ZWang/etal:08}.
For the simple cases where each chain only has two spins, it was
determined that perfect entanglement transfer can be realized between
spin pairs by adjusting the magnetic field strength and the anisotropy
parameter.

In this paper, we study quantum communication in an anisotropic Heisenberg
XY chain with small number of particles $N$ and an arbitrary initial state.
Following the scenario of earlier work on state transfer through spin
chains, we encode a state to be transferred at the first site of the chain
and, without external control, let the chain freely evolve. After time
$t$ the state is to be readout at the $r^\text{th}$ site of the chain. But
different from the other cases with $[\sum\nolimits_{i}S_{i}^{z},H]=0$, the
quantum communication channel, the spin chain, not only transfers the state,
but also generates entanglement.

This paper is organized as follows: In part two we will give our model
Hamiltonian and calculate the time dependence of the one-site correlation
function which will be used in the expression of the fidelity and
tangle. In part three, we discuss the time evolution of transmission
fidelity and purity
in a short chain for three regimes: strong-field, weak-field and
intermediate regime. The final part is devoted to the conclusions.


\section{The model and the calculation}

The Hamiltonian of the anisotropic Heisenberg XY chain in a uniform
transverse magnetic field $h$ is given by \cite{Anderson:58}%
\begin{equation}
H=-\sum%
\limits_{i=1}^{N}(J^{x}S_{i}^{x}S_{i+1}^{x}+J^{y}S_{i}^{y}S_{i+1}^{y})-\sum%
\limits_{i=1}^{N}hS_{i}^{z}.  \label{eq:anisH}
\end{equation}%
where $S_{i}^{a}$ is the $a$ spin operator ($a=x,y,z$) at site $i $, $J^{x}$
and $J^{y}$ are the anisotropic exchange interaction constants, and $h$ is
the transverse magnetic field. We assume periodic boundary conditions, so
that the $N^{\text{th}}$ site is identified with the $0^{\text{th}}$ site.
The standard procedure used to solve Eq.~(\ref{eq:anisH}) is to transform
the spin operators $S_{i}^{a}$ into fermionic operators via the
Jordan-Wigner(J-W) transformation
\begin{equation}
c_{1}=S_{1}^{-},c_{n}=(-2S_{1}^{z})(-2S_{2}^{z})\cdot \cdot \cdot
(-2S_{n-1}^{z})S_{n}^{-},n=2,...N,
\end{equation}%
where $c_{n}$ are one-dimensional spinless fermions annihilation operators
and $S_{1}^{-}=$ $S_{1}^{x}-iS_{1}^{y}$.

It will be convenient to introduce operators $A_{l}=c_{l}^{\dagger }+c_{l},$
and $B_{l}=c_{l}^{\dagger }-c_{l},$ which fulfill the anti-commutation
relations
\begin{equation}
\{A_{l},A_{m}\}=-\{B_{l},B_{m}\}=2\delta _{lm},\;\;\;\{A_{l},B_{m}\}=0,
\end{equation}%
In terms of these operators, the J-W transformation reads
\begin{equation}
S_{l}^{x}=\frac{1}{2}A_{l}\prod\limits_{s=1}^{l-1}A_{s}B_{s},\;\;S_{l}^{y}=-%
\frac{i}{2}B_{l}\prod\limits_{s=1}^{l-1}A_{s}B_{s},\;\;S_{l}^{z}=-\frac{1}{2}%
A_{l}B_{l},  \label{eq:AnB}
\end{equation}%
Using the J-W transformation, Eq.~(\ref{eq:anisH}) becomes the bi-linear
form
\begin{equation}
H=-\sum_{i}\left\{ \frac{J}{2}\left[ \left( c_{i}^{\dagger }c_{i+1}-c_{i}c_{i+1}^{\dagger }\right) +\gamma \left( c_{i}^{\dagger }c_{i+1}^{\dagger
}-c_{i}c_{i+1}\right) \right] +h\left( c_{i}^{\dagger }c_{i}-\frac{1}{2}\right) \right\}.  \label{eq:Hc}
\end{equation}%
where $J=\frac{1}{2}(J^{x}+J^{y}),$ and $\gamma =(J^{x}-J^{y})/(J^{x}+J^{y})$
is the anisotropy parameter. The limiting values, $\gamma =0$ and $1$
correspond to the isotropic and Ising chain, respectively.

Eq.~(\ref{eq:Hc}) can be diagonalized by the transformation \cite%
{Amico/etal:04,LAmico/etal:04}
\begin{equation}
\eta _{k}=\frac{1}{\sqrt{N}}\sum_{l}e^{ikl}(\alpha _{k}c_{l}+i\beta
_{k}c_{l}^{\dagger }),  \label{eq:eta}
\end{equation}%
where
\begin{eqnarray}
\alpha _{k} &=&\sqrt{\frac{1-(h+J\cos k)/\lambda _{k}}{2}},  \notag \\
\beta _{k} &=&sign(J\gamma \sin k)\sqrt{\frac{1+(h+J\cos k)/\lambda _{k}}{2}}%
,  \notag \\
\lambda _{k} &=&\sqrt{(h+J\cos k)^{2}+J^{2}\gamma ^{2}\sin ^{2}k}, \label{eq:Afa}
\end{eqnarray}%
The Hamiltonian in Eq.~(\ref{eq:Hc}) then becomes
\begin{equation}
H=\sum_{k}\lambda _{k}\left(\eta _{k}^{\dagger }\eta _{k}-\frac{1}{2}\right).
\end{equation}%
where the wave number $k=2\pi m/N$ with $-N/2<m\leq N/2.$

Now that the \textit{XY} Hamiltonian has been diagonalized, we will
calculate the evolution of the operators $c_{j}^{\dagger },c_{j}$, which
will be used to obtain the final state. From Eq.~(\ref{eq:eta}) and its
inverse
\begin{equation}  \label{eq:coft}
c_{j}(t)=\frac{1}{\sqrt{N}}\sum_{k}e^{-ikj}(\alpha _{k}\eta _{k}(t)-i\beta
_{k}\eta _{-k}^{\dagger }(t)),
\end{equation}
where $\eta _{k}(t)=e^{-i\lambda _{k}t}$ $\eta _{k}(0)$.

Then from Eqs.~(\ref{eq:eta}) and (\ref{eq:coft})
\begin{eqnarray}  \label{eq:cs}
c_{j}(t) &=&\sum_{l}[\overset{\thicksim }{a}_{lj}(t)c_{l}+\overset{\thicksim
}{b}_{lj}(t)c_{l}^{\dagger }],  \notag \\
c_{j}^{\dagger }(t) &=&\sum_{l}[\overset{\thicksim }{b}_{lj}^{\dagger
}(t)c_{l}+\overset{\thicksim }{a}_{lj}^{\dagger }(t)c_{l}^{\dagger }],
\end{eqnarray}
where
\begin{eqnarray}  \label{eq:coeffsc}
\overset{\thicksim }{a}_{lj}(t)&=&\frac{1}{N}\sum_{k}e^{ik(l-j)}[e^{i\lambda
_{k}t}-2i\alpha _{k}^{2}\sin \lambda _{k}t],  \notag \\
\overset{\thicksim }{b}_{lj}(t)&=&\frac{2}{N}\sum_{k}e^{ik(l-j)}\alpha
_{k}\beta _{k}\sin \lambda_{k}t.
\end{eqnarray}
For $\gamma =0$, and $\alpha _{k}=0,$ the evolution of the creation operator
is
\begin{equation*}
c_{j}^{\dagger }(t)=\frac{1}{N}\sum_{l}e^{-ik(l-j)}e^{-i\lambda
_{k}t}c_{l}^{\dagger }.
\end{equation*}
In this case the $z$-component of the total spin $S^{z}$ commutes with the
Hamiltonian and is a conserved quantity.

Now we seek to transfer a quantum state from the site $s$ to $r$. First we
assume that all the spins of the system are initially in spin down states.
Then we encode the state $\left\vert \varphi (0)\right\rangle =\alpha
\left\vert 0\right\rangle +\beta \left\vert 1\right\rangle $ at the first
spin of the chain. The initial state of the whole system is then $\left\vert
\Phi (0)\right\rangle =(\alpha +\beta c_{1}^{\dagger })\left\vert \mathbf{0}%
\right\rangle$, where $\left\vert \mathbf{0}\right\rangle $ denotes the
state with all the spins down. For simplicity, $\alpha$ and $\beta $ are
taken to be real with $\alpha ^{2}+\beta ^{2}=1$. Now our task is to
calculate the fidelity of transmission of a state from the first spin to the
$r^\text{th}$ spin of the chain.

The reduced density matrix of the $r^\text{th}$ spin of the chain can be
constructed using the operator expansion for the density matrix of a system
on $N$ spin-1/2 particles in terms of tensor products of Pauli matrices. The
reduced density matrix is \cite{Osborne/etal:02}
\begin{equation}  \label{eq:dmat}
\rho _{r}(t)=\left(
\begin{array}{cc}
\frac{1}{2}+\left\langle S_{r}^{z}\right\rangle & \left\langle
S_{r}^{x}\right\rangle -i\left\langle S_{r}^{y}\right\rangle \\
\left\langle S_{r}^{x}\right\rangle +i\left\langle S_{r}^{y}\right\rangle &
\frac{1}{2}-\left\langle S_{r}^{z}\right\rangle%
\end{array}%
\right).
\end{equation}
where $\left\langle S_{r}^{x,y,z}\right\rangle $ means $\left\langle \Phi
(0)\right\vert S_{r}^{x,y,z}\left\vert \Phi (0)\right\rangle .$

The fidelity between the received state $\rho _{r}(t)$ and the initial state
$\varphi(0)$ is defined by $F=\sqrt{\left\langle \varphi (0)\right\vert \rho
_{r}(t)\left\vert \varphi (0)\right\rangle }$ which is
\begin{equation}  \label{eq:f}
F=\sqrt{\frac{1}{2}+(\beta ^{2}-\alpha ^{2})\left\langle
S_{r}^{z}\right\rangle +2\alpha \beta \left\langle S_{r}^{x}\right\rangle}.
\end{equation}
So if the parameters $\alpha =\beta =1/\sqrt{2}$, the fidelity $F=\sqrt{%
\frac{1}{2}+\left\langle S_{r}^{x}\right\rangle }$, which only depends on
the value of $\left\langle S_{r}^{x}\right\rangle $.

Another quantity we want to calculate is the purity (also known as the tangle or one-tangle), which provides a measure of the entanglement between the
spin at one site and the rest of sites in the chain. The purity is often expressed as $1-\tr(\rho^2)$, but can also be expressed as
\begin{equation}  \label{eq:tangle}
\tau[\rho ^{(1)}]=4\det [\rho ^{(1)}],
\end{equation}
where $\rho ^{(1)}$ is the one-site reduced density matrix, Eq.~(\ref%
{eq:dmat}). This is referred to as the one-tangle
\cite{LAmico/etal:04}, which was apparently motivated by the tangle
defined in Ref.~\cite{Coffman/Kundu/Wootters}.
Note that Eq.~(\ref{eq:tangle}) is gives a valid measure of
entanglement when the whole
system is in a pure state and given there is only one parameter for a
any such measure for a two-state system, this is as good as any
other. However, if the system is in a mixed state, one must use
some other measure of entanglement.  The one-tangle, or purity,
is connected to the Von Neumann entropy of the reduced density matrix
through the relation
\begin{equation}
S[\rho ^{(1)}]=h(\frac{1}{2}(1+\sqrt{1-\tau \lbrack \rho ^{(1)}]})),
\end{equation}
where $h(x)=-x\log _{2}x-(1-x)\log _{2}(1-x)$. From Eq.~(\ref{eq:dmat}), the
tangle/purity can be written as
\begin{equation}  \label{eq:tao}
\tau \lbrack \rho ^{(1)}]=4\det [\rho ^{(1)}]=1-4(\left\langle
S_{r}^{x}\right\rangle ^{2}+\left\langle S_{r}^{y}\right\rangle
^{2}+\left\langle S_{r}^{z}\right\rangle ^{2}),
\end{equation}
where the components $S^{\alpha}$ are the components of the Bloch
vector.

Now we will calculate $\left\langle S_{r}^{a}(t)\right\rangle $ ($a=x,y,z$)
for use in Eqs.~(\ref{eq:f}) and ~(\ref{eq:tao}) .
\begin{equation}
\left\langle S_{r}^{x}(t)\right\rangle =\alpha ^{2}\left\langle \mathbf{0}%
\right\vert S_{r}^{x}(t)\left\vert \mathbf{0}\right\rangle +\beta
^{2}\left\langle \mathbf{1}\right\vert S_{r}^{x}(t)\left\vert \mathbf{1}%
\right\rangle +\alpha \beta (\left\langle \mathbf{0}\right\vert
S_{r}^{x}(t)\left\vert \mathbf{1}\right\rangle +\left\langle \mathbf{1}%
\right\vert S_{r}^{x}(t)\left\vert \mathbf{0}\right\rangle).\label{eq:sxx}
\end{equation}
However, note that $\left\langle \mathbf{0}\right\vert
S_{r}^{x}(t)\left\vert \mathbf{0}\right\rangle =\left\langle \mathbf{1}%
\right\vert S_{r}^{x}(t)\left\vert \mathbf{1}\right\rangle =0$.

Using Wick's theorem,
\begin{eqnarray}
\left\langle S_{r}^{x}(t)\right\rangle &=&\frac{1}{2}\alpha \beta
\{\left\langle \mathbf{0}\right\vert (A_{1}c_{1}^{\dag
}+c_{1}A_{1})\left\vert \mathbf{0}\right\rangle \left\langle \mathbf{0}%
\right\vert B_{1}A_{2}B_{2}...A_{r-1}B_{r-1}A_{r}\left\vert \mathbf{0}%
\right\rangle -  \notag \\
&&\left\langle \mathbf{0}\right\vert (B_{1}c_{1}^{\dag
}+c_{1}B_{1})\left\vert \mathbf{0}\right\rangle \left\langle \mathbf{0}%
\right\vert A_{1}A_{2}B_{2}...A_{r-1}B_{r-1}A_{r}\left\vert \mathbf{0}%
\right\rangle +  \notag \\
&&\cdot \cdot \cdot +\left\langle \mathbf{0}\right\vert (A_{r}c_{1}^{\dag
}+c_{1}A_{r})\left\vert \mathbf{0}\right\rangle \left\langle \mathbf{0}%
\right\vert A_{1}B_{1}A_{2}B_{2}...A_{r-1}B_{r-1}\left\vert \mathbf{0}%
\right\rangle \}.
\end{eqnarray}

For the simple case, $N=3$, $r=2$%
\begin{eqnarray}
\left\langle S_{r}^{x}(t)\right\rangle &=&\frac{\alpha \beta }{2}%
\{\left\langle \mathbf{0}\right\vert
(A_{1}c_{1}^{\dagger}+c_{1}A_{1})\left\vert \mathbf{0}\right\rangle
\left\langle \mathbf{0}\right\vert B_{1}A_{2}\left\vert \mathbf{0}%
\right\rangle  \notag \\
&&\;\;\;\; -\left\langle \mathbf{0}\right\vert (B_{1}c_{1}^{\dag
}+c_{1}B_{1})\left\vert \mathbf{0}\right\rangle \left\langle \mathbf{0}%
\right\vert A_{1}A_{2}\left\vert \mathbf{0}\right\rangle  \notag \\
&&\;\;\;\; +\left\langle \mathbf{0}\right\vert (A_{2}c_{1}^{\dag
}+c_{1}A_{2})\left\vert \mathbf{0}\right\rangle \left\langle \mathbf{0}%
\right\vert A_{1}B_{1}\left\vert \mathbf{0}\right\rangle.
\end{eqnarray}
The value of $\left\langle S_{r}^{y}\right\rangle $ can be obtained from the
above expression by replacing $A_{r}\rightarrow B_{r}$ and $1/2\rightarrow
-i/2$.

Now we need to calculate \textit{contractions} of two field operators $%
\left\langle \mathbf{0}\right\vert A_{j}B_{m}\left\vert \mathbf{0}%
\right\rangle$, $\left\langle \mathbf{0}\right\vert A_{j}A_{m}\left\vert
\mathbf{0}\right\rangle$, $\left\langle \mathbf{0}\right\vert
B_{j}B_{m}\left\vert \mathbf{0}\right\rangle$, $\left\langle \mathbf{0}%
\right\vert B_{j}A_{m}\left\vert \mathbf{0}\right\rangle$. From Eqs.~(\ref%
{eq:cs})- (\ref{eq:coeffsc}), we get
\begin{eqnarray}  \label{eq:sx}
\left\langle \mathbf{0}\right\vert A_{j}(t)B_{m}(t)\left\vert \mathbf{0}%
\right\rangle &=&\delta _{jm}  \notag \\
&&-\frac{4}{N}\sum_{k}[2\alpha _{k}^{2}\beta _{k}^{2}\cos k(j-m)  \notag \\
&&+\alpha _{k}\beta _{k}(1-2\beta _{k}^{2})\sin k(j-m)]\sin ^{2}\lambda _{k}t
\notag \\
&&+\frac{2}{N}\sum_{k}\alpha _{k}\beta _{k}\cos k(j-m)\sin 2\lambda _{k}t, \\
\left\langle \mathbf{0}\right\vert A_{j}(t)A_{m}(t)\left\vert \mathbf{0}%
\right\rangle &=&\delta _{jm}  \notag \\
&&-\frac{4i}{N}\sum_{k}[\alpha _{k}\beta _{k}(1-2\alpha _{k}^{2})\cos k(j-m)
\notag \\
&&-2\alpha _{k}^{2}\beta _{k}^{2}\sin k(j-m)]\sin ^{2}\lambda _{k}t  \notag
\\
&&+\frac{2i}{N}\sum_{k}\alpha _{k}\beta _{k}\sin k(j-m)\sin 2\lambda _{k}t,
\\
\left\langle \mathbf{0}\right\vert B_{j}(t)B_{m}(t)\left\vert \mathbf{0}
\right\rangle &=&-\delta _{jm}  \notag \\
&&-\frac{4i}{N}\sum_{k}[\alpha _{k}\beta _{k}(1-2\alpha _{k}^{2})\cos k(j-m)
\notag \\
&&+2\alpha_{k}^{2}\beta _{k}^{2}\sin k(j-m)]\sin ^{2}\lambda _{k}t  \notag \\
&&+\frac{2i}{N}\sum_{k}\alpha _{k}\beta _{k}\sin k(j-m)\sin 2\lambda _{k}t.\label{eq:sxbb}
\end{eqnarray}
and
\begin{eqnarray}
\left\langle \mathbf{0}\right\vert A_{j}(t)c_{1}^{\dag
}+c_{1}A_{j}(t)\left\vert \mathbf{0}\right\rangle &=&\frac{2}{N}%
\sum_{k}[\cos k(1-j)\cos \lambda _{k}t  \notag \\
&&-(1-2\alpha _{k}^{2})\sin k(1-j)\sin \lambda _{k}t  \notag \\
&&+2\alpha _{k}\beta _{k}\cos k(1-j)\sin \lambda _{k}t], \\
\left\langle \mathbf{0}\right\vert B_{j}(t)c_{1}^{\dag
}+c_{1}B_{j}(t)\left\vert \mathbf{0}\right\rangle &=&\frac{-2i}{N}%
\sum_{k}[(1-2\alpha _{k}^{2})\cos k(1-j)\sin \lambda _{k}t  \notag \\
&&+\sin k(1-j)\cos \lambda _{k}t  \notag \\
&&+2\alpha _{k}\beta _{k}\sin k(1-j)\sin \lambda _{k}t].
\end{eqnarray}

Now we can calculate $\left\langle S_{r}^{z}\right\rangle $%
\begin{equation}  \label{eq:sz}
\left\langle S_{r}^{z}\right\rangle =-\frac{1}{2}[\alpha ^{2}\left\langle
\mathbf{0}\right\vert A_{r}B_{r}\left\vert \mathbf{0}\right\rangle +\beta
^{2}\left\langle \mathbf{1}\right\vert A_{r}B_{r}\left\vert \mathbf{1}%
\right\rangle +\alpha \beta (\left\langle \mathbf{0}\right\vert
A_{r}B_{r}\left\vert \mathbf{1}\right\rangle +\left\langle \mathbf{1}%
\right\vert A_{r}B_{r}\left\vert \mathbf{0}\right\rangle )].
\end{equation}
Note that, since $\left\langle \mathbf{0}\right\vert A_{r}B_{r}\left\vert
\mathbf{1}\right\rangle \mathbf{=}\left\langle \mathbf{1}\right\vert
A_{r}B_{r}\left\vert \mathbf{0}\right\rangle =0,$ then $\left\langle \mathbf{%
0}\right\vert A_{r}(t)B_{r}(t)\left\vert \mathbf{1}\right\rangle$ = \newline
$\left\langle \mathbf{1}\right\vert A_{r}(t)B_{r}(t)\left\vert \mathbf{0}%
\right\rangle$ = $0$ \cite{LAmico/etal:04},
\begin{equation}
\left\langle S_{r}^{z} \right\rangle =-\frac{1}{2}\left[\left\langle \mathbf{%
0}\right\vert A_{r}B_{r}\left\vert \mathbf{0}\right\rangle + \beta
^{2}\left( |\overset{\thicksim}{b}_{1r}|^{2}- |\overset{\thicksim }{a}%
_{1r}|^{2}\right)\right].
\end{equation}
Now, from Eq.~(\ref{eq:sx}), $\left\langle \mathbf{0}\right\vert
A_{r}B_{r}\left\vert \mathbf{0}\right\rangle =1-\frac{8}{N}\sum_{k}\alpha
_{k}^{2}\beta _{k}^{2}\sin ^{2}\lambda _{k}t.$

For the case $\gamma =0$, $\alpha _{k}=0$, we find that
\begin{equation}
\left\langle S_{r}^{x}(t)\right\rangle =\frac{\alpha \beta }{N}\sum_{k}\cos
[k(r-1)-\lambda _{k}t],\;\;\;\;\;\; \left\langle S_{r}^{z}\right\rangle
=\beta ^{2}\left\vert \overset{\thicksim }{a}_{1r}\right\vert ^{2}-\frac{1}{2%
}.
\end{equation}
which agrees with Bose's case \cite{Bose:03}. The $\overset{\thicksim }{a}%
_{1r}$ gives the transmission amplitude $f_{rs}^{N}(t)$ of an excitation
(the $\left\vert 1\right\rangle $ state) from the $1^\text{st}$ to the $r^%
\text{th} $ spin. When $\alpha =\beta =1/\sqrt{2}$, the fidelity is
\begin{equation}
F=\sqrt{\frac{1}{2}+\frac{1}{2}\frac{1}{N}\sum_{k}\cos [k(r-1)-\lambda _{k}t]%
}.
\end{equation}
which agrees with Ref.~\cite{Wang/etal:08}.


\section{Results and discussion:  Fidelity of state transfer}

We will now investigate the performance of a spin chain for which there exists some anisotropy parameter $\gamma$ and the $z$-component of the total
spin is not a conserved quantity. We first perform numerical calculations of the fidelity $F$ for a particular, and small $N$ value and discuss the
variations with changes in physical properties of the system.

As stated above, after some time $t$, the state at the $r^\text{th}$
spin will be measured, where $r=N/2+1$ for $N$ even and $r=(N+1)/2+1$
for $N$ odd), i.e. the sending and receiving positions are at opposite sites on
the chain with periodic boundary conditions. Clearly the complexity of the
computation grows with the value of $r$ because the expansion of $%
\left\langle S_{r}^{x}(t)\right\rangle $ has $(2r-1)!!$ terms. So for simplicity, we take $N=5, r=3$ as an example. We will also choose a particular
state ($\sqrt{3}/2\left\vert 0\right\rangle+1/2\left\vert 1\right\rangle$) to analyze. However, it is important to note that we have examined several
states and have found these trends typical; they exhibit similar variations, only the maximal values of the fidelity and purity/tangle are different. In
each case the final state Eq.~(\ref{eq:dmat}) has a time dependence described by Eqs.~(\ref{eq:sxx}) and (\ref{eq:sz}). The state to be transferred is
encoded at the first site of the chain. Then as time evolves the state propagates to the $r^\text{th}$ site. The time evolution of the operators
$\left\langle S_{r}^{a}(t)\right\rangle (a=x,y,z)$ can be expressed as a product of all contractions of the two operators $A(t)$ and $B(t)$, which are
superpositions of many different states in the basis spanned by the Jordan-Wigner fermions with definite momenta \cite{Crooks/etal:08}. This is clearly
more complex than the isotropic case ($\gamma =0$) where the state transfer can be characterized in terms of their dispersion \cite{Bose:03}.


\subsection{The strong-field regime}

\begin{figure}[tbp]
\begin{center}
\includegraphics[scale=0.55,angle=0]{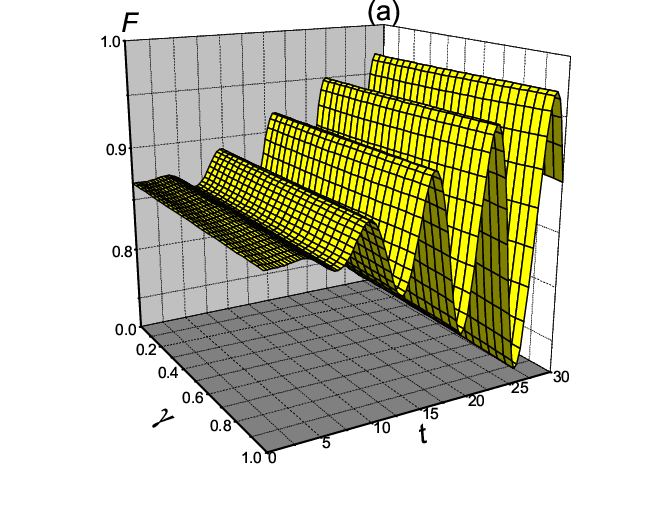} %
\includegraphics[scale=0.55,angle=0]{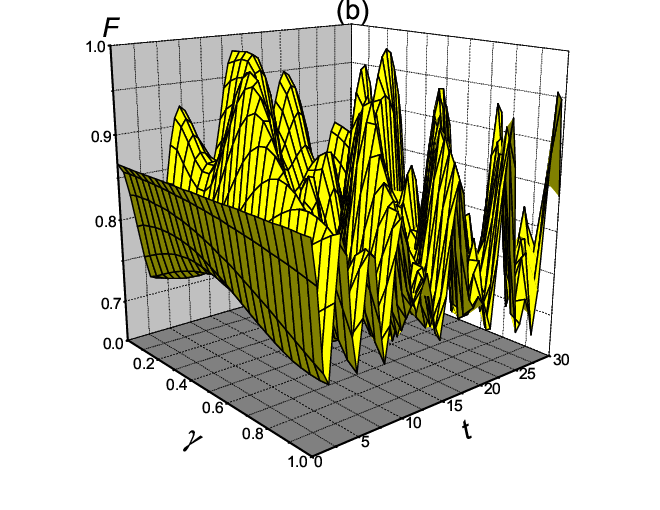} %
\includegraphics[scale=0.55,angle=0]{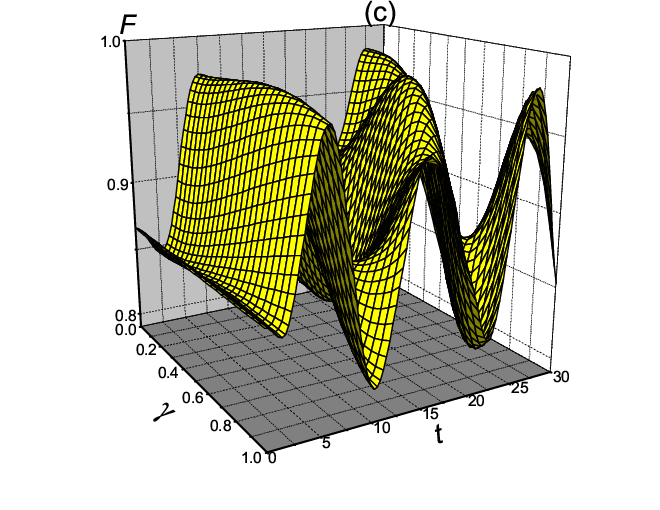}
\end{center}
\caption{The fidelity $F$ is plotted versus time $t$ and anisotropic
  parameter $\protect\gamma$ for (a) Strong-field regime, $h=1.0,$
  $J=0.1$. (b) Weak-field regime, $h=0.1,J=1.0$.(c) Intermediate case,
  $h=J=0.5$. The number of sites $N$ is 5, the input state is encoded
  in the first site and output state is at the $3^\text{rd}$ site,
  $\protect\alpha =\protect\sqrt{3}/2$,$\protect\beta =1/2$.} \label{fig:1}
\end{figure}

The fidelity $F$ as a function of time $t$ and the anisotropic parameter $%
\gamma $ are shown in Fig.~\ref{fig:1} with different parameter values for $%
J $ and $h$. First in the strong-field regime (Fig.~\ref{fig:1}(a)), $J/h\ll 1$, the dispersion is given by $\lambda _{k}=h[1+(J/h)\cos k+(J/h)^{2}(\cos
^{2}k+\gamma ^{2}\sin ^{2}k)/2+o((J/h)^{3})]$. The effect of the anisotropy can be neglected if we only consider the term which is first order in $J/h$.
The parameters in Eq.~(\ref{eq:Afa}) become $\alpha _{k}\rightarrow 0,\beta _{k}\rightarrow \pm 1$ and $\overset{\thicksim }{b}_{lj}\rightarrow 0$ in
Eq.~(\ref{eq:coeffsc}).  So the contractions of two field operators in Eqs.~(\ref{eq:sx})-~(\ref{eq:sxbb}) are negligible for $j\neq m$ which indicates
that using only a uniform magnetic field (global interaction) cannot cause a uncorrelated state to become correlated. In Fig.~\ref{fig:1}(a), it is
clear that the anisotropy does not significantly affect the fidelity, but the presence of the cosine term in the first order of the dispersion produces
the observed oscillation of fidelity with time $t$.


\subsection{The weak-field regime}

In the weak-field regime $h/J\ll 1$ (Fig.~\ref{fig:1}(b)), the
dispersion can be written as $\lambda _{k}=J[2(h/J)\cos
k+(h/J)^{2}+\cos ^{2}k+\gamma ^{2}\sin ^{2}k]^{1/2}$.
In this case, the anisotropy has pronounced effects on the
fidelity. We see that increasing the anisotropy does not always
decrease the fidelity.  For certain values of the parameters $\gamma $
and $t$, the fidelity is greater, which corresponds to a constructive
interference.  For example, there are peaks with $\gamma \neq 0$ such
as ($\gamma =0.28,t=27.70,F=0.98$) and ($\gamma =0.42,t=7.10,F=0.98$)
that exhibit this behavior.  With increasing $\gamma $ and $t$, the
frequency of oscillation of $F$ becomes greater.  The higher-frequency
oscillation can be attributed to the fact that with increasing $\gamma
$ the initial state differs more from the true ground state, and
therefore, with the evolution of time, it exhibits fluctuations, even
near the ground state.


\subsection{The intermediate regime}

The intermediate regime $(J\sim h)$, which is the most interesting, is
shown in Fig.~\ref{fig:1}(c).  For relatively short times $0<t<10,$
the anisotropy does not have a pronounced effect on the fidelity. The
oscillation of $F$ with time $t$ is relatively slow in this case
compared to the cases in both the strong- and weak-field regimes,
Fig.~\ref{fig:1}(a) and (b) respectively. When the strength of the
exchange interaction $J$ or the magnetic field $h$ dominate, the
oscillation of $F$ with time $t$ is greater. But for intermediate
regimes, the competition between $J$ and $h$ gives a smaller 
oscillation frequency for $F$.  Furthermore, the high-fidelity peaks
are fairly broad and therefore correspond to values of the fidelity
which are stable under perturbation of the parameters.


\subsection{Comparison}

In summary, realistic quantum communication devices will require
larger fidelities in a shorter times.  One may well have expected a
significant oscillation of fidelity with time and anisotropy in the
intermediate regime.  However, this is not the case, but rather the
intermediate regime has the highest fidelities in the shortest amount
of time with the more stable values for the fidelity.  So if one wants
to achieve the highest fidelity under our assumed conditions, 
the intermediate regime is, surprisingly, the best choice.


\section{Results and discussion:  Entanglement}

\begin{figure}[tbp]
\begin{center}
\includegraphics[scale=0.55,angle=0]{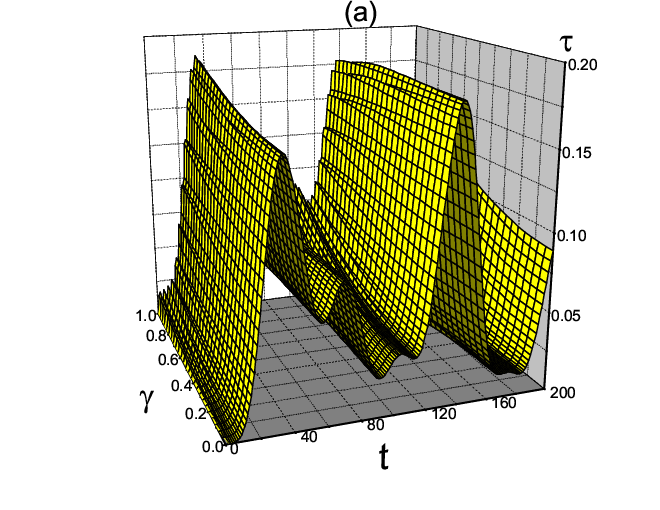} %
\includegraphics[scale=0.55,angle=0]{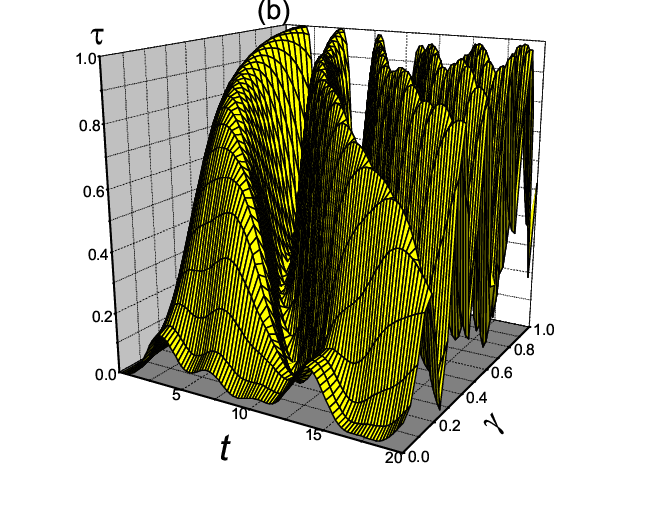} %
\includegraphics[scale=0.55,angle=0]{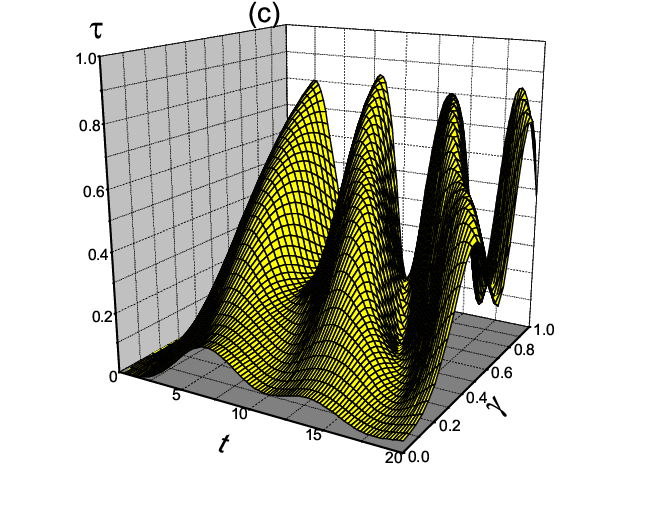}
\end{center}
\caption{The one-tangle, or purity, at the $3^{\text{rd}}$ site as a function of time $t$
and $\protect\gamma $. The initial state of the system is $(\protect\sqrt{3}%
/2\left\vert 0\right\rangle +1/2\left\vert 1\right\rangle )\otimes \left\vert \mathbf{0}\right\rangle $. (a) Strong-field regime, $h=1.0, J=0.1$.
 (b) Weak-field regime, $h=0.1,J=1.0$.
  (c) Intermediate regime, $h=J=0.5$. } \label{fig:2}
\end{figure}

\begin{figure}[htbp]
\begin{center}
\includegraphics[scale=0.55,angle=0]{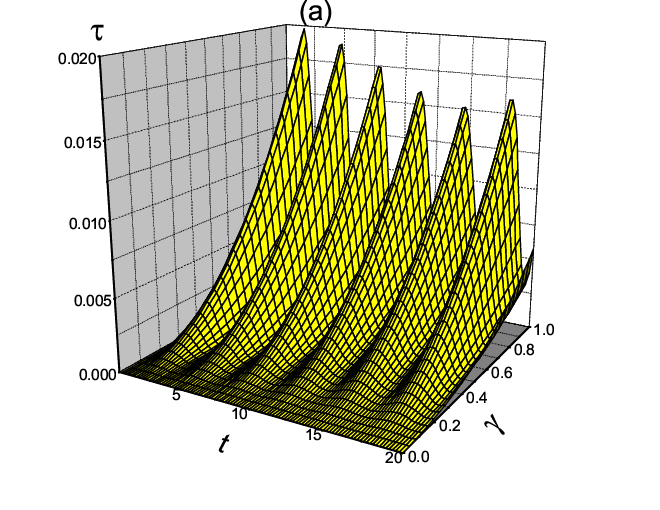} %
\includegraphics[scale=0.55,angle=0]{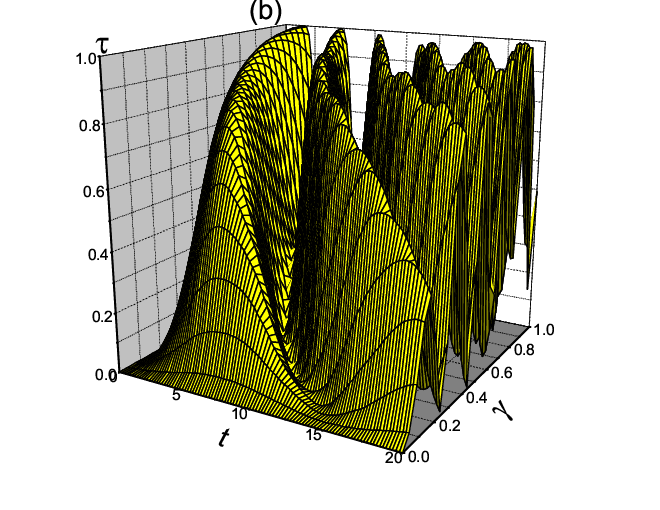} %
\includegraphics[scale=0.55,angle=0]{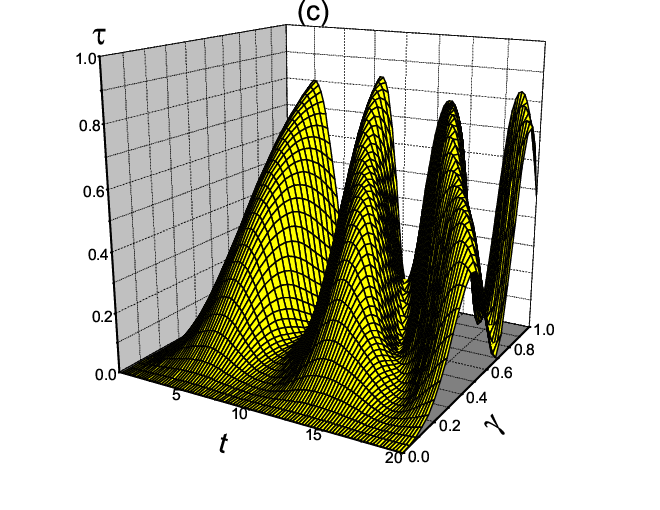}
\end{center}
\caption{Same as Fig.~\protect\ref{fig:2} except that the initial state is
the vacuum state, $\left\vert \mathbf{0}\right\rangle$.}
\label{fig:3}
\end{figure}

It is clear that the initial state of the system is an unentangled state.
However, with the evolution of time, entanglement is generated in the chain.
In order to quantify this change, we have calculated the purity, or
tangle, at the
$3^{\text{rd}}$ site which measures the entanglement between the $3^{\text{rd}%
}$ site and all the other sites in the chain. The tangle at $3^{\text{rd}%
}$ site as a function of time $t$ and anisotropy $\gamma $ is plotted in Fig.~\ref{fig:2} and Fig.~\ref{fig:3} with different initial states $(\alpha
\left\vert 0\right\rangle +\beta \left\vert 1\right\rangle )\otimes \left\vert \mathbf{0}\right\rangle $ and vacuum state, respectively. From Fig.
\ref{fig:2}, for small anisotropy ($\gamma <0.1$), the tangle is negligibly small. In the strong-field regime, from Fig. \ref{fig:3}(a), the anisotropy
does not have a significant affect on the tangle similar to behaviour of the fidelity as seen in Fig. \ref{fig:2}(a).  Also, the strong-field tangle is
relatively small compared to the weak-field regime and the oscillation of tangle with time $t$ is suppressed. Furthermore, the time for the tangle reach
its first peak in the strong-field regime is at $t\approx 45$ while for the intermediate regime it is $t\approx 5$.

In the weak-field regime, which is plotted in Fig.~\ref{fig:2}(b) and
Fig.~\ref{fig:3}(b), we see that a stronger exchange interaction even
with a stronger anisotropy can generate more entanglement at certain
times. For example, at $\gamma =1.0$ and $t=1.5$, $\tau =1.0$.

In the intermediate regime, the behaviour of the tangle is similar to the weak-field regime.  With increasing anisotropy, the tangle increases. However,
even in the $\gamma =1.0$ case, it does not reach its maximal value $\tau =1.0$. Also, as $J/h$ increases, the tangle oscillates more rapidly, but
different from the effect of anisotropy on fidelity in the intermediate case, (see Fig. \ref{fig:2}(b)) the tangle increases with increasing $\gamma$.
This is an effect of entanglement dynamically generated from the
ground state and depends on the anisotropy. When the initial state is the ground state, the
system will still generate entanglement, which exists only in the
$\gamma \neq 0$ case. This is due to the double spin-flip operator terms
$\gamma(c_{i}^{\dagger }c_{i+1}^{\dagger }-c_{i}c_{i+1})$ in
Eq.~(\ref{eq:Hc}).  In this case, 
$\left\langle \mathbf{0}\right\vert S_{r}^{x}(t)\left\vert\mathbf{0}\right\rangle
=\left\langle \mathbf{0}\right\vert S_{r}^{y}(t)\left\vert \mathbf{0}%
\right\rangle =0$, and $\tau \lbrack \rho ^{(1)}]=1-4\left\langle \mathbf{0}%
\right\vert S_{r}^{z}(t)\left\vert \mathbf{0}\right\rangle ^{2}$. Unlike
Fig.~\ref{fig:2}, the one-tangle always equals zero when $\gamma =0$ since
the vacuum state is the ground state of the system. And the $r^{\text{th}}$
spin will always remain in the spin-down state and will never be entangled
with the other spins. For the two initial states of the system, Fig.~\ref%
{fig:2}(b)(c) and Fig.~\ref{fig:3}(b)(c), we find that the time evolution of
the one-tangle shows similar behavior with increasing $\gamma $, which means
the one-tangle is not sensitive to the initial state of the system when a
strong anisotropy is present in the case of weak-field and intermediate
regimes.  This shows that the anisotropy parameter aides in the
generation of entanglement in the spin chain.

\section{Conclusions}

In conclusion, we have investigated quantum state transfer through an anisotropic
 Heisenberg XY model in a transverse field and also entanglement
generation in that same system. The interest in these problems
stems from the possible use of spin chains as a communication
 channels. We expect that a
realistic ferromagnetic material would have some anisotropy and
 have shown that this anisotropy can have significant effects on the fidelity of state
transfer as well as the entanglement in the chain.

Specifically, we have calculated the fidelity and the one-tangle, or
purity, for three different cases: a weak external magnetic field, an
intermediate regime, and strong external magnetic field. We found that
in all three cases a relatively high fidelity can be obtained for
certain times and vales of the anisotropy. However, in the intermediate
regime, the oscillation of the fidelity with time is fairly low, and
the peaks fairly broad.  Furthermore, a fairly high fidelity is
achieved in a relatively shorter time for the intermediate regime.  
This would imply a more reliable output, in a shorter time, when
some anisotropy is present, compared to when there is none. Thus the
intermediate regime presents some interesting and somewhat surprising
results for state transfer and is also the best choice for reliable
transfer.  

We have also calculated the one-tangle, or the purity of a typical
one-particle state in the chain.  
We began with a pure initial state and found that the stronger the anisotropy
and exchange interaction, the more entanglement will be generated.
This indicates that anisotropy also aides in the production of
entanglement in the chain.  


\section*{ACKNOWLEDGMENTS}

This material is based upon work supported by the National Science
Foundation under Grant No. 0545798 to MSB. ZMW thanks the scholarship
awarded by the China Scholarship Council(CSC). We gratefully acknowledge C.
Allen Bishop for helpful discussions.




\end{document}